%% file: sens_standalone.tex
\documentclass{article}
\usepackage{graphicx}
\usepackage{dcolumn}
\usepackage{bm}
\usepackage{color}  
\definecolor{orange}{rgb}{1,0,0}

\usepackage{amsmath}

\usepackage[lofdepth,lotdepth]{subfig}

\title{Sterile Neutrino Sensitivity with Wrong-Sign Muon Appearance at $\nu$STORM}
\author{C.D. Tunnell}
\date{}                                           

\begin{document}

\maketitle

\vspace*{-6.cm}
\begin{flushright}
IDS-NF-035
\end{flushright}


\include{chap_sens/sensitivity}

\section*{Acknowledgements}  The author thanks Alan Bross, John Cobb, and Joachim Kopp for their guidance and knowledge.  The author also thanks Ryan Bayes for the migration matrices used in this analysis.

\bibliography{refs}        
\bibliographystyle{plain}  
\end{document}

%% file: chap_sens/sensitivity.tex
\newcommand{\mymomentum}{3.8}
\newcommand{\mymomentumspread}{0.38}
\newcommand{\mybaseline}{2000 }


\begin{figure}[t]	
\begin{center}
\includegraphics[width=0.7\columnwidth]{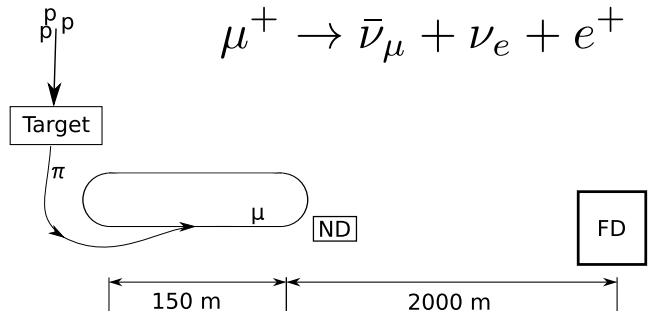}
\end{center}
\vspace*{-0.5cm}
\caption{\label{fig:facility}A diagram of the proposed $\nu$STORM facility.}
\end{figure}

Neutrinos from STORed Muons\footnote{The facility was previously called the Very Low Energy Neutrino Factory (VLENF).} ($\nu$STORM) is a proposed experiment that uses 3.8 GeV/c muon decay to produce a well-understood beam of electron and muon neutrinos that can be used for short baseline physics (Fig.~\ref{fig:facility}).  A magnetized far detector allows for the wrong-sign muon appearance physics of $\nu_e \to \nu_\mu$ and provides more sensitivity to sterile neutrinos than other proposals (See comparisons in \cite{Abazajian:2012ys}) because of the relative ease to which muon tracks can be identified.  Other physics such as $\nu_e$ and $\nu_\mu$ cross section measurements are possible.  For further details, see the Letter of Intent \cite{c:nustorm_loi}.  

An explanation of the $\nu$STORM appearance analysis will follow. This work is a continuation of the work presented in \cite{Tunnell:2011ya}.  For disappearance measurement work, see \cite{Winter:2012sk}.

\section{Short Baseline Oscillations}

LEP experiments revealed that there are three light neutrinos that couple to the $Z$-boson (ie. \emph{active neutrinos}), however, there are theoretical and experimental motivations \cite{Abazajian:2012ys} for neutrinos without Standard Model interactions called \emph{sterile} neutrinos.  The (3+1) scenario is the case of three active neutrinos with an additional heavy sterile neutrino -- $m_4 >> m_\text{others}$ -- and only this situation is considered although the results are generalizable. 

The probability $\nu_e \to \nu_\mu$ depends on the mixing matrix $U$ (Reviewed in \cite{c:pdg}).  Let $R_{ij}$ be a rotation between the $i$-th and $j$-th mass eigenstates without a CP violating phase: CP violation cannot be observed in oscillations with large $\Delta m^2$ dominance (See p.g. 273 of \cite{0198508719}).  For $N$ neutrinos, $R_{ij}$ has dimension $N \times N$ and takes the form:

\begin{eqnarray}
R_{ij} = \begin{pmatrix}
1 & \ldots & 0 & \ldots & 0 & \ldots & 0\\
\vdots &  & \vdots &  & \vdots &  & \vdots\\
0 & \ldots & \cos \theta_{ij} & \ldots &  \sin \theta_{ij} & \ldots & 0\\
\vdots &  & \vdots &  & \vdots &  & \vdots\\
0 & \ldots & - \sin \theta_{ij} & \ldots &  \cos \theta_{ij} & \ldots & 0\\
\vdots & & \vdots &  & \vdots &  & \vdots\\
0 & \ldots & 0 & \ldots & 0 & \ldots & 1\\
\end{pmatrix}.
\end{eqnarray}

By convention, the three neutrino mixing matrix is $U_\text{PMNS} = R_{23} R_{13} R_{12}$. In the (3+1) model of neutrino oscillations, extra rotations can be introduced such that the mixing matrix is $U_\text{(3+1)} = R_{34} R_{24} R_{14} U_\text{PMNS}$.  Given that $\Delta m^2_{41} >> \Delta m^2_{31}$, $U_\text{PMNS}$ can be approximated by the identity matrix (\emph{ie.} the ``short baseline approximation") implying $U_{e 4} = \sin(\theta_{14})$ and $ U_{\mu 4} = \sin(\theta_{24}) \cos(\theta_{14}) $.

The oscillation probabilities for appearance and disappearance, respectively, are:
\begin{eqnarray}
\label{eq:prob}\text{P}_{\nu_e \to \nu_\mu} &=& 4 | U_{e 4}|^2 |U_{\mu 4}|^2 \sin^2 \left(\frac{\Delta m^2_{41} L}{4 E}\right)\\
&=&\sin^2 (2 \theta_{e\mu})\sin^2 \left(\frac{\Delta m^2_{41} L}{4 E}\right),\\
\label{eq:probdisp}\text{P}_{\nu_\alpha \to \nu_\alpha} &=& 1 - \left[4 |U_{\alpha 4}|^2 (1 - |U_{\alpha 4}|^2)\right] \sin^2 \left(\frac{\Delta m^2_{41} L}{4 E}\right).
\end{eqnarray}

\noindent
in this short baseline limit where the definition $\sin^2 (2 \theta_{e\mu}) =  4 | U_{e 4}|^2 |U_{\mu 4}|^2$ has been introduced.

Electron and muon neutrino disappearance measurements will constrain $| U_{e 4}|^2$ (\cite{Winter:2012sk}) and  $| U_{\mu 4}|^2$ while the appearance channel analysis could measure the product $|U_{e 4}|^2 |U_{\mu 4}|^2$.   Information about the matrix element $U_{e 4}$ also arises from jointly analyzing $\bar{\nu}_\mu$ disappearance and $\nu_\mu$ appearance.  The remaining matrix element $U_{\tau 4}$ can be extracted by analyzing NC rates $|U_{s 4}|^2 = \sum_{e,\mu,\tau} |U_{\alpha 4}|^2$, using the other channels to constrain $| U_{e 4}|^2$ and $| U_{\mu 4}|^2$, and assuming unitarity.

	\section{The Neutrino Flux: $\Phi$}

\begin{table}[t]
\centering
  \caption{Matrix elements for muon decay}
  \begin{tabular}{ r || c | c  }
	& $f_0(x)$ & $f_1(x)$\\
	\hline \hline
	$\nu_\mu$ & $2 x^2 (3 - 2 x)$ &  $2 x^2 (1 - 2x)$\\
	\hline
	$\nu_e$ & $12 x^2 (1 - x)$ &  $12 x^2 (1 - x)$\\
  \end{tabular}
    \label{t:matrix_elements}
\end{table}

Muon-decay beams contrast pion-decay beams because the beam characteristics and production mechanisms are well-known.  The neutrino flux arises from the Electroweak decay of $\mu \to \nu_\mu \bar{\nu}_e e$ and it is sufficient to compute matrix elements at tree level.  The neutrino spectrum for a $\mu^\pm \to e^\pm + \nu_e (\bar{\nu}_e) + \bar{\nu_\mu} (\nu_\mu)$ decay in the rest frame of the muon follows:

\begin{eqnarray}
\label{l:rate}
\frac{\operatorname{d} n}{\operatorname{d}x \operatorname{d}\Omega} = \frac{1}{4 \pi} \left[ f_0(x) \mp \mathcal{P} f_1(x) \cos\theta \right]
\end{eqnarray}

\noindent
where $x = 2 E^\text{c.o.m.}_\nu / m_\mu \in [0,1]$ is the scaled neutrino energy in the rest frame, $\Omega$ is the solid angle in the rest frame, $f_0(x)$ and $f_1(x)$ are muon decay parameters, and $\mathcal{P}$ is the polarization.  Electron and neutrino masses are negligible for this process and ignored, hence the inclusive range for values of $x$.  These muon decay paramters can be computed to leading order with Electroweak theory (See, for example, chapter 6 of Ref.~\cite{0521266033}) and are neutrino flavor dependent (See Table~\ref{t:matrix_elements}).  

\begin{figure}[t]	
\begin{center}
\includegraphics[width=0.7\columnwidth]{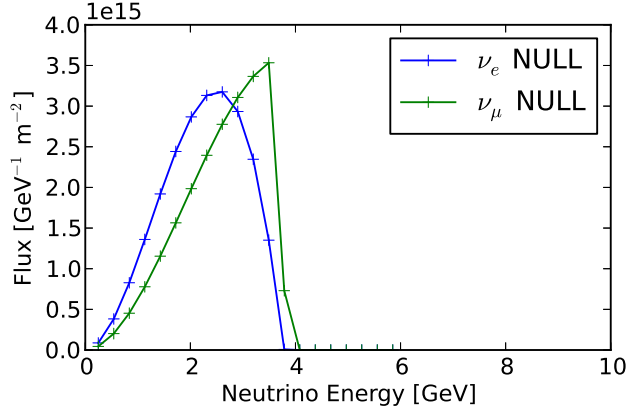}
\end{center}
\vspace*{-0.5cm}
\caption{\label{fig:simple_flux}The flux of $\nu_e$ and $\bar{\nu}_\mu$ for a \mymomentum GeV/c muon-decay without oscillations at \mybaseline  meters.  No smearing due to accelerator effects has been performed.}
\end{figure}

The polarization $\mathcal{P}$ is set to zero, similar to other studies, and has been shown to average to zero due to Thomas Precession.  Boosting the neutrino distributions into the lab frame leads to:

\begin{eqnarray}
\label{eq:boost_flux}
\frac{\operatorname{d}^2 N_\mu}{\operatorname{d}y \operatorname{d}A} &=&  \frac{4 n_\mu}{\pi L^2 m^6_\mu} E^4_\mu y^2 (1 - \beta \cos \phi) \left[3 m^2_\mu - 4 E^2_\mu y (1 - \beta \cos\phi)\right]\\
\frac{\operatorname{d}^2 N_e}{\operatorname{d}y \operatorname{d}A} &=&  \frac{24 n_\mu}{\pi L^2 m^6_\mu} E^4_\mu y^2 (1 - \beta \cos \phi) \left[m^2_\mu - 2 E^2_\mu y (1 - \beta \cos\phi)\right]
\end{eqnarray}

\noindent
where $y = E_\nu / E_\mu$ is the scaled neutrino energy in the lab frame, $\beta = \sqrt{1 - m^2_\mu / E^2_\mu}$, $A$ is an area, and $n_\mu$ is the number of muons. These neutrino distributions (Fig.~\ref{fig:simple_flux}) are for a point source so they are not directly applicable to the decay straight of $\nu$STORM.

The number of muons assumed is $1.8 \times 10^{18}$ and is based on $10^{21}$ protons on target (POT) at 60 GeV/c.  It corresponds to roughly 5 years of running with a 100 kW target station.  The number of useful muon decays is motivated in \cite{c:nustorm_loi}.

\begin{figure}[t]	
\begin{center}
\includegraphics[width=0.7\columnwidth]{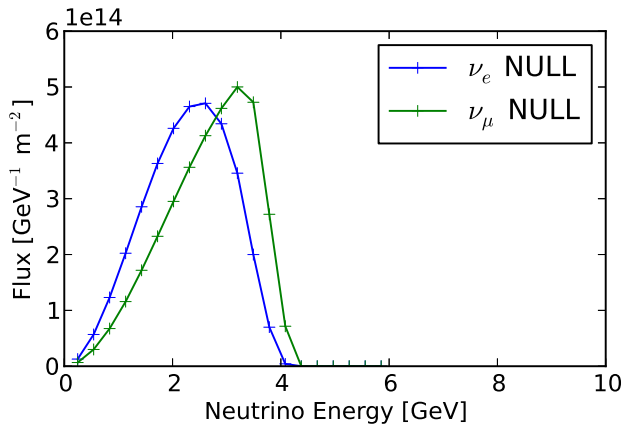}
\end{center}
\vspace*{-0.5cm}
\caption{\label{fig:nooscflux}The unoscillated flux of $\nu_e$ and $\bar{\nu}_\mu$ for a $(\mymomentum \pm \mymomentumspread) \text{ GeV/c}$ muon-decay at \mybaseline  meters.  Accelerator effects are included; see the text for details.}
\end{figure}

\begin{figure}[t]	
\begin{center}
\includegraphics[width=0.7\columnwidth]{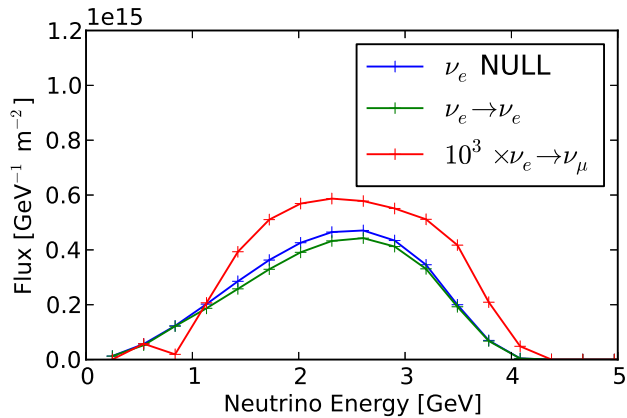}
\end{center}
\vspace*{-0.5cm}
\caption{\label{fig:lsndflux}The flux at the far detector for a $(\mymomentum \pm \mymomentumspread) \text{ GeV/c}$  muon for initial $\nu_e$ states including integration over the beam envelope and detector volume.  Final states include $\nu_e$ without oscillations and both $\nu_e$ and $\bar{\nu}_\mu$ with best fit short baseline oscillations.  The normalization is $10^{21}$ POT.}
\end{figure}

When computing the flux for $\nu$STORM, the far detector approximation of a point-source accelerator and detector no longer is applicable since the size of the detector and accelerator straight (150 meters) are comparable to the baseline of 2000 meters. The neutrino fluxes are computed by integrating over the decay straight, transverse beam phase space, and detector volume.   The beam occupies a 6D phase space ($x$, $y$, $z$, $p_x$, $p_y$, $p_z$) and the detector has a $5\text{ m} \times 5\text{ m}$ cross section with the depth set by the desired fiducial mass of 1.3 kt.  Both transverse 2D phase spaces are represented by the Twiss parameters $\alpha = 0$ and $\beta = 40 \text{ m}$ where the $1 \sigma$ Gaussian geometric emittance is assumed to be $2.1 \text{ mm}$.  The spread in, for example, $x$ is $\sigma_x = \sqrt{\beta \epsilon}$ and the angular divergence in $x$ is $\sigma_{x'} = \sqrt{\epsilon / \beta}$. The longitudinal phase space ($z$ and $p_z$) is described by assuming a uniform distribution in $z \in [0, 150 \text{ m}]$ -- accurate to 0.5\% -- and $p_z \in [3.8 \pm \mymomentumspread \text{ GeV/c}]$. 

The flux is computed by Monte Carlo (MC) integration: random points are chosen within the beam phase space and within the detector volume to determine the expected flux.  This integration introduces a new computational requirement:  the baseline is a variable that affects both the oscillation probability ($L/E$) and the flux ($L^{-2}$ geometric factor).  The GLoBES software (version 3.1.10)  \cite{Huber:2004ka,Huber:2007ji} that is used for neutrino factory phenomenology treats these as separable problems and was modified to compute this flux (and later the event rates and sensitivities).  Specifically, GLoBES is modified such that both the flux and oscillation probability are computed in the \emph{oscillation probability engine}.   The code for the analysis is available \cite{vlenf_tools} under the GPL license \cite{gpl}.  

The resulting flux after the integration (Fig.~\ref{fig:nooscflux}) is corrected for accelerator effects.  The corrections are small for far detector physics (Compare to Fig.~\ref{fig:simple_flux}) but are important for near detector physics where the baseline is smaller than the decay straight.

\section{The Oscillation Probability: $\text{(Prob.)}$}
	
\begin{figure}[t]	
\begin{center}
\includegraphics[width=0.7\columnwidth]{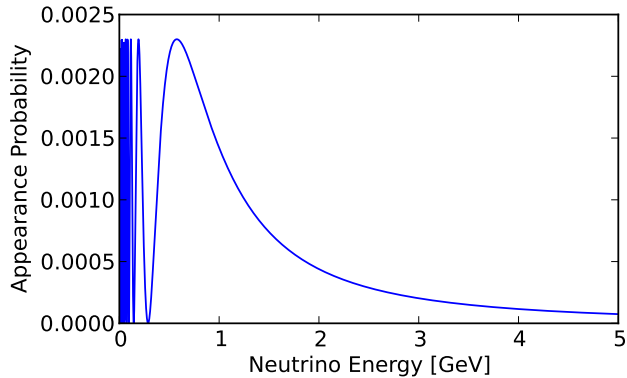}
\end{center}
\vspace*{-0.5cm}
\caption{\label{fig:osc_prob}The oscillation probability for the ``golden channel" $\nu_e \to \nu_\mu$ from Eq.~\ref{eq:prob} using the (3+1) oscillation parameters in TABLE \ref{tab:params}.  A baseline of \mybaseline meters is assumed.}
\end{figure}
	
	This section will discuss how sterile oscillation phenomenology relates to conducting the proposed experiment.  For instance, for a point-source baseline of \mybaseline meters, it is possible to determine the oscillation probability (Fig.~\ref{fig:osc_prob}) using Eq.~\ref{eq:prob} for any combination of $L$ and $E$. 

\begin{table}[b]
\centering
\caption{\label{tab:params}%
Best-fit oscillation parameters for the (3+1) sterile neutrino scenario using combined MB $\bar{\nu}$ and LSND $\bar{\nu}$ data \cite{Giunti:2011hn}.}
\begin{tabular}{|l|r|}
\hline
\textrm{Parameter}&
\textrm{Value}\\
\hline
$\Delta m^2_{41}$ [$\text{eV}^2$] & 0.89\\
$|U_{e4}|^2$ & 0.025\\
$|U_{\mu 4}|^2$ & 0.023\\
\hline
\end{tabular}
\end{table}

\begin{table}[t]  
\centering
  \caption{Values for $3\times3$ oscillations used.}
  \begin{tabular}{ |c |}
  \hline
$\sin^2\theta_{12} = 0.319$ \\
$\sin^2\theta_{23} = 0.462 $\\
$\sin^2\theta_{13} = 0.010 $\\
$\Delta m^2_{21} = 7.59 \times 10^{-5} \text{ eV}^2$ \\
$\Delta m^2_{31} = 2.46 \times 10^{-3} \text{ eV}^2$ \\
\hline
  \end{tabular}
  \label{t:pmnsfit}
\end{table}

The best fit parameters for the ``short baseline anomaly'' and $3\times3$ mixing (\emph{i.e.} $\sin^2 (2 \theta_{13})$, $\Delta m^2_{12}$, etc.) are used throughout the analysis.  The best fit parameters for the LSND anomaly come from  \cite{Giunti:2011hn} (See TABLE~\ref{tab:params}) and agree with those published by the LSND collaboration \cite{PhysRevLett.81.1774}.  For completeness, oscillations between known mass eigenstates are included despite not influencing the sensitivity: the correction is order  $10^{-5}$.  The best fit data from \cite{GonzalezGarcia:2010er} is used to specify standard $3\times3$ oscillations.   Without loss of generality, normal hierarchy is assumed and the values of known $3\times3$ mixing can be seen in Table~\ref{t:pmnsfit}.  Errors associated with these quantities are ignored.

Computationally, the SNU (version 1.1) add-on  \cite{Kopp:2006wp,Kopp:2007ne} has been used to extend computations in GLoBES to $4\times4$ mixing matrices.

\section{Cross section: $\sigma$}

\begin{figure}
  \centering
  \subfloat[CC]{\label{fig:xseccc}\includegraphics[width=0.5\textwidth]{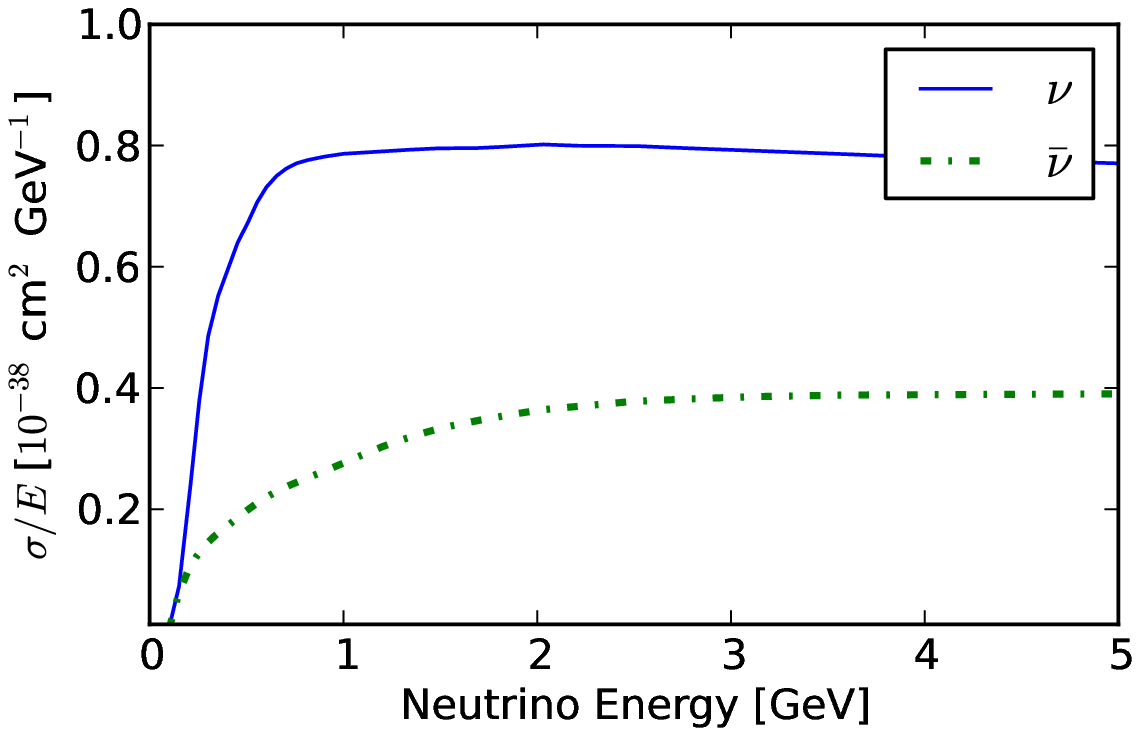}}                
  \subfloat[NC]{\label{fig:xsecnc}\includegraphics[width=0.5\textwidth]{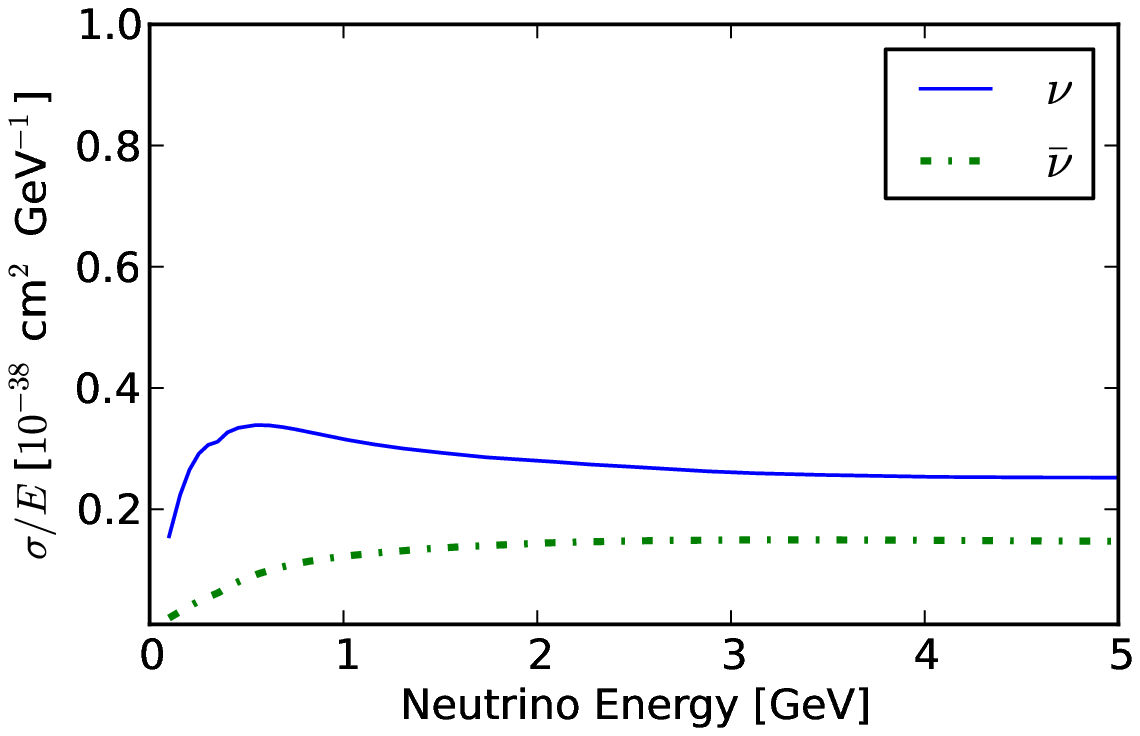}}
  \caption{Neutrino cross sections per nucleon.}
  \label{fig:xsec}
\end{figure}

Cross sections are required for each neutrino flavor ($\nu_\mu$, $\bar{\nu}_\mu$, $\nu_e$, $\bar{\nu}_e$) and each interaction type (CC or NC).  The nucleon cross sections (Fig.~\ref{fig:xsec}) are calculated in \cite{Messier:1999kj} and \cite{Paschos:2001np} for the low energy and high energies, respectively.   NC cross sections are flavor independent.  The CC cross sections are approximately flavor independent: \emph{Fermi's Second Golden Rule} results in the same matrix elements and, at these energies, the phase spaces for the final-state electrons and muons are equal.

The total cross section requires knowing the number of nucleons in addition to the nucleon cross section.  The fiducial mass of 1.3 kt determines the number of nucleons via Avogadro's number.

\section{Interaction rates: $N_\text{int.}$}

\begin{figure}
  \centering
  \subfloat[Appearance with stored $\mu^+$]{\label{fig:gull}\includegraphics[width=0.5\textwidth]{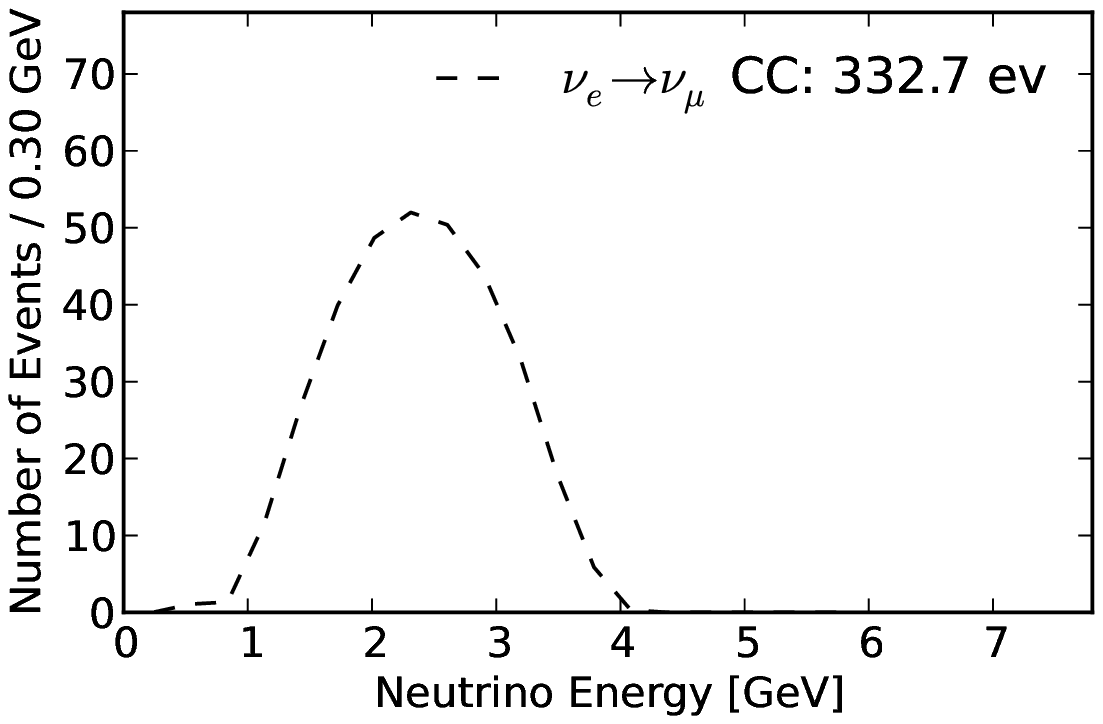}}                
  \subfloat[Appearance with stored $\mu^-$]{\label{fig:tiger}\includegraphics[width=0.5\textwidth]{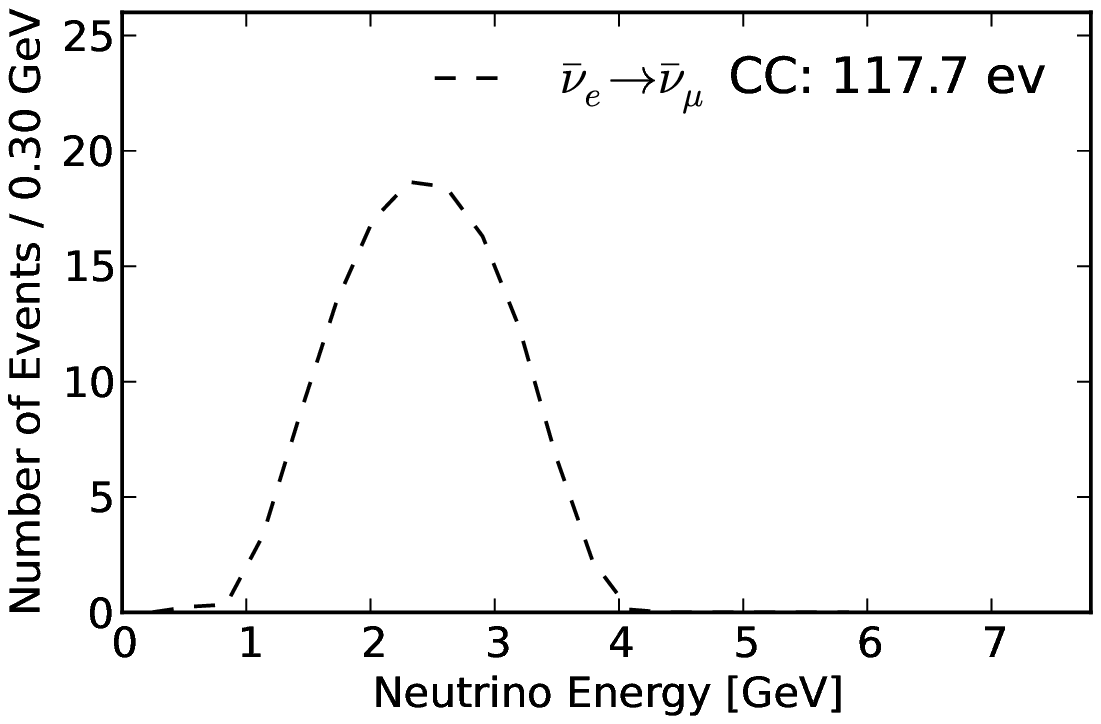}}\\
  \subfloat[Disappearance with stored $\mu^+$]{\label{fig:mouse}\includegraphics[width=0.5\textwidth]{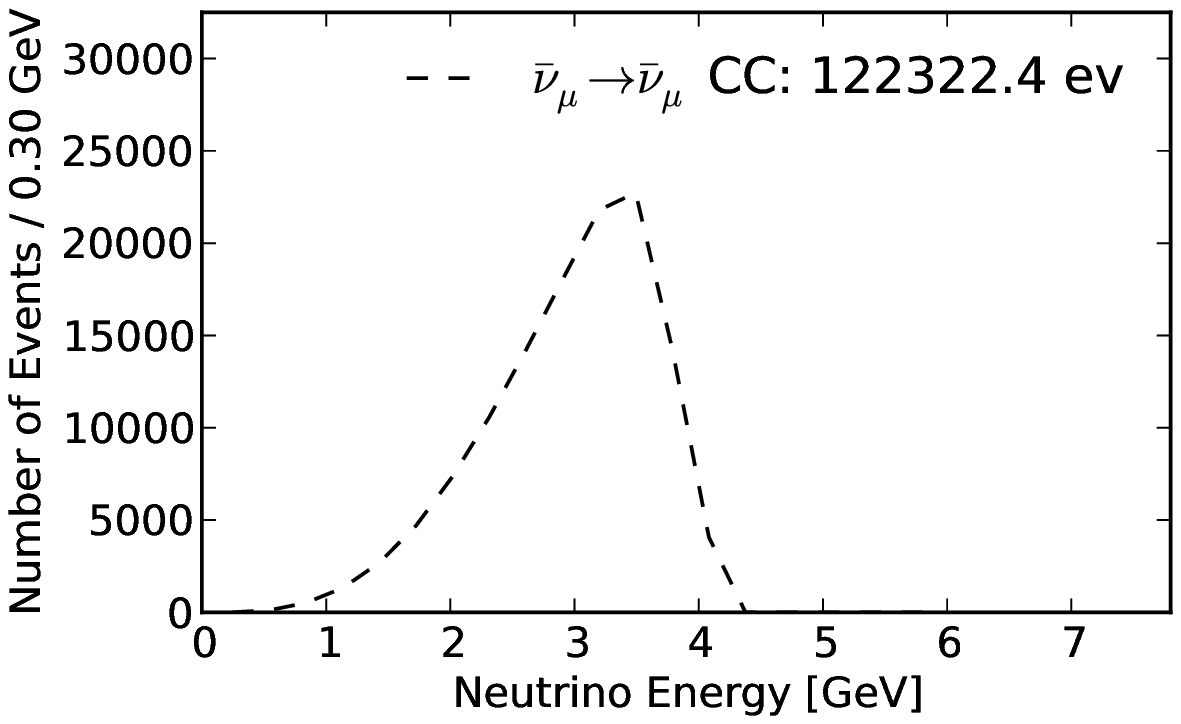}}
   \subfloat[Disappearance with stored $\mu^-$]{\label{fig:mouse2}\includegraphics[width=0.5\textwidth]{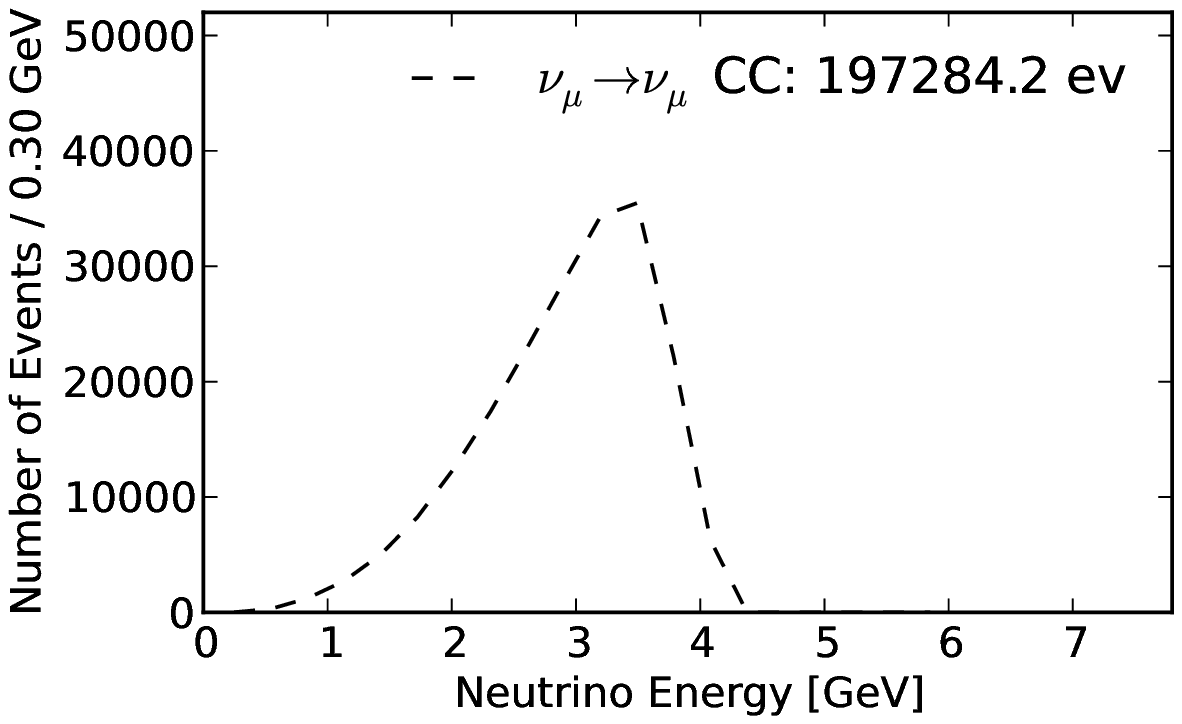}}
  \caption{True channel rate energy distributions assuming the LSND anomaly best fit values.  The transitions $\nu_e \to \nu_\mu$, $\bar{\nu}_e \to \bar{\nu}_\mu$, $\bar{\nu}_\mu \to \bar{\nu}_\mu$, and $\nu_\mu \to \nu_\mu$ are shown.  }
  \label{fig:channel_rates}
\end{figure}

The number of neutrino interactions is computed which does not require assumptions about the detector. The interaction rates can be computed by $N_\text{int.} = \Phi \times \text{(Prob.)}  \times \sigma$, for flux $\Phi$, oscillation probability $(\text{Prob.})$, and cross section $\sigma$, where all of these quantities have been computed in the previous sections.  

Using the LSND anomaly best fit (TABLE~\ref{tab:params}) as a example for a sterile neutrino signal, the event rates for $\mu^+$ and $\mu^-$ decays are shown in TABLE~\ref{tab:raw_evt}.  Various deductions can be made about these event rates and their statistical significance.  With either stored $\mu^+$s or stored $\mu^-$s, the statistical significance of all channels is greater than $10\sigma$.  Combining the NC channels together results in a statistical significance of $20\sigma$ and $17\sigma$ for stored $\mu^+$ and $\mu^-$, respectively.  There are no known physics backgrounds to neither $\nu_e \to \nu_mu$ CC nor $\bar{nu}_e \to \bar{\nu}_mu$ CC interactions except to negligible solar-term oscillations, so the backgrounds will arise from how well the detector can differentiate these interactions.

The number of events can also be determined as a function of energy since the evolution of $\rho$, $\sigma$, and $(\text{Prob.})$ as a function of energy is known.  These distributions are shown in Fig.~\ref{fig:channel_rates}.

There are numerous channels with reach into the sterile neutrino parameter space.  Most other experiments have one channel to explore (See \cite{Abazajian:2012ys} for list of experiments), whereas in the best case $\nu$STORM allows for 10 signals and in the worst case 6 (i.e. combine $\nu_e \to \nu_e$ CC and all NC channels).

\begin{table}[h]
	\centering
	\caption{\label{tab:raw_evt}Truth event rates for $10^{21}$ POT for the no oscillations and short baseline oscillations described by TABLE~\ref{tab:params}.  The statistical significances are computed.  The combined statistical significance of NC events are 20 and 17 for stored $\mu^+$ and $\mu^-$, respectively.  There are no physics backgrounds to $\nu_e \to \nu_mu$ CC interactions.}
	\subfloat[Stored $\mu^-$.]{
\begin{tabular}{c|r|r|r|r}
Channel & $N_\textrm{osc.}$ & $N_\textrm{null}$ & Diff. & $(N_\textrm{osc.} - N_\textrm{null})/\sqrt{N_\textrm{null}}$\\
        \hline
$\bar{\nu}_e \to \bar{\nu}_\mu$ CC & 117 & 0 & $\infty$ & $\infty$ \\
$\bar{\nu}_e \to \bar{\nu}_e$ NC & 30511 & 32481 & -6.1\% & -10.9\\
$\nu_\mu \to \nu_\mu$ NC & 66037 & 69420 & -4.9\% & -12.8\\
$\bar{\nu}_e \to \bar{\nu}_e$ CC & 77600 & 82589 & -6.0\% & -17.4\\
$\nu_\mu \to \nu_\mu$ CC & 197284 & 207274 & -4.8\% & -21.9\\
\end{tabular}
	}	
	\qquad

	\subfloat[Stored $\mu^+$.]{
\begin{tabular}{c|r|r|r|r}
Channel & $N_\textrm{osc.}$ & $N_\textrm{null}$ & Diff. & $(N_\textrm{osc.} - N_\textrm{null})/\sqrt{N_\textrm{null}}$\\
        \hline
$\nu_e \to \nu_\mu$ CC & 332 & 0 & $\infty$ & $\infty$ \\
$\bar{\nu}_\mu \to \bar{\nu}_\mu$ NC & 47679 & 50073 & -4.8\% & -10.7\\
$\nu_e \to \nu_e$ NC & 73941 & 78805 & -6.2\% & -17.3\\
$\bar{\nu}_\mu \to \bar{\nu}_\mu$ CC & 122322 & 128433 & -4.8\% & -17.1\\
$\nu_e \to \nu_e$ CC & 216657 & 230766 & -6.1\% & -29.4\\
\end{tabular}
	}
\end{table}

\section{Event rates after cuts}

It must be determined how many of the raw events pass analysis cuts. Similar analyses have been performed for Neutrino Factories exploring CP violation at energies ranging from 25 GeV \cite{NF:2011aa} to 5 GeV \cite{Geer:2007kn}, but never at 3.8 GeV/c.  Preexisting experience and knowledge exists as to fractional background levels and analysis difficulties; work had to be performed in order to tune the analysis for this energy range.

The detector performance can be represented by \emph{migration matrices} (also known as response matrices or energy smearing matrices) that describe both the energy resolution and detection efficiency.   If events are binned in terms of true neutrino energy then the migration matrix is needed to transform the distribution into the space of measured neutrino energies.  For example, take the histogram:

\begin{equation}
\vec{h}^\text{true} = (N^\text{true}_{\text{0.0 - 0.1 \text{GeV}}}, N^\text{true}_{\text{0.1 - 0.2 \text{GeV}}}, \ldots, N^\text{true}_{\text{3.9 - 4.0 \text{GeV}}})^T, 
\end{equation}

\noindent
where $N^\text{true}_{\text{0.0 - 0.1 \text{GeV}}}$ is the number of events in the bin with ranges 0.0 and 0.1 GeV.  The migration matrix $\mathbf{M}$ used for this analysis is a square matrix and defined such that $\vec{h}^\text{measured} = \mathbf{M} \vec{h}^\text{true}$ where $\vec{h}^\text{measured}$ is the expected histogram of reconstructed quantities in the detector.

With a perfect detector $ \mathbf{M} = \text{diag.}( 1, 1, \ldots, 1)$.  $\mathbf{M}$ is unitary if and only if it describes only energy smearing.  Efficiencies are included into $\mathbf{M}$ by removing the unitarity constraint.  

\begin{figure}
  \centering
  \subfloat[$\nu_\mu$ CC]{\label{fig:mm_app}\includegraphics[width=0.5\textwidth]{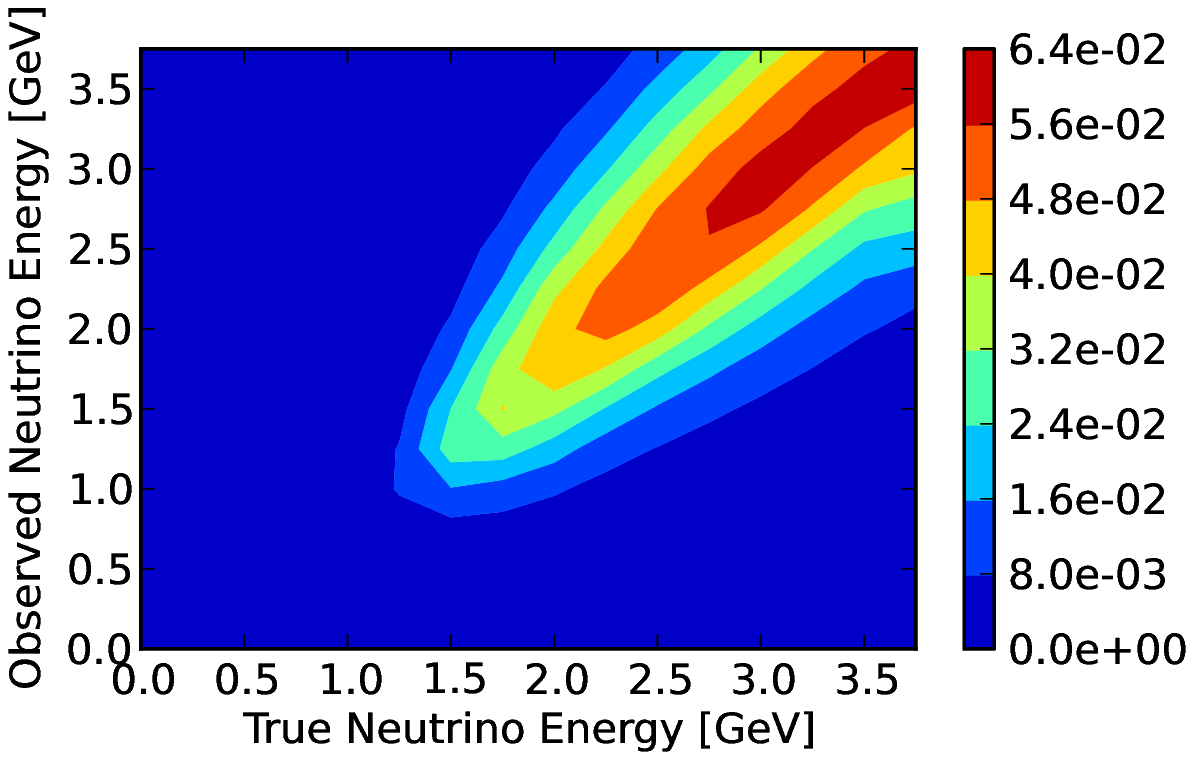}}                
  \subfloat[$\bar{\nu}$ NC]{\label{fig:mm_mubarnc}\includegraphics[width=0.5\textwidth]{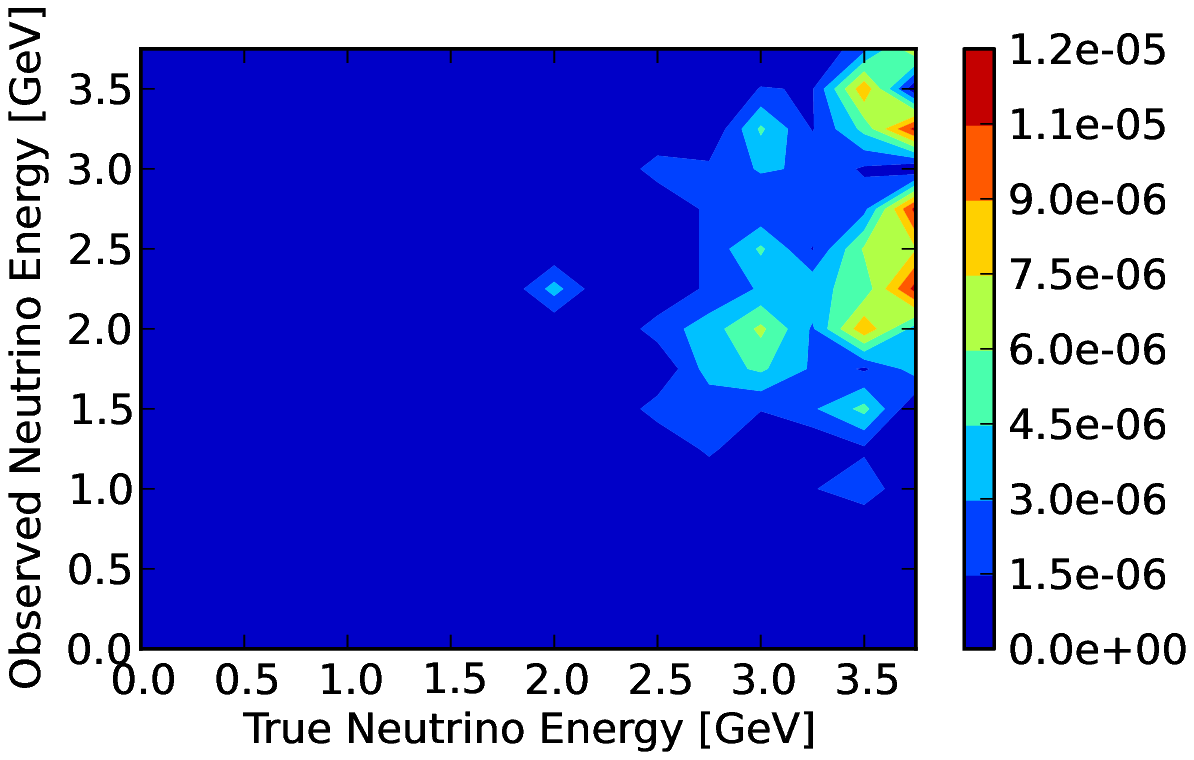}}\\
  \subfloat[$\bar{\nu}_\mu$ CC]{\label{fig:mubarcc}\includegraphics[width=0.5\textwidth]{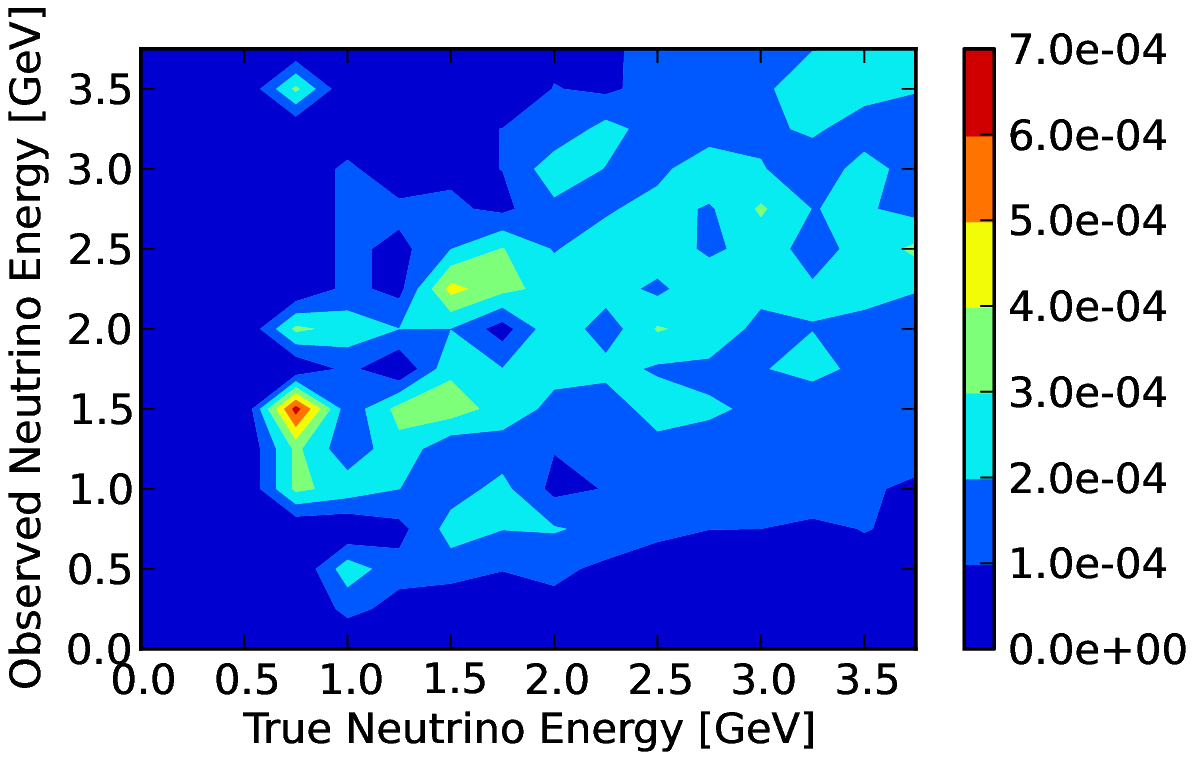}}
   \subfloat[$\nu_e$ CC]{\label{fig:ecc}\includegraphics[width=0.5\textwidth]{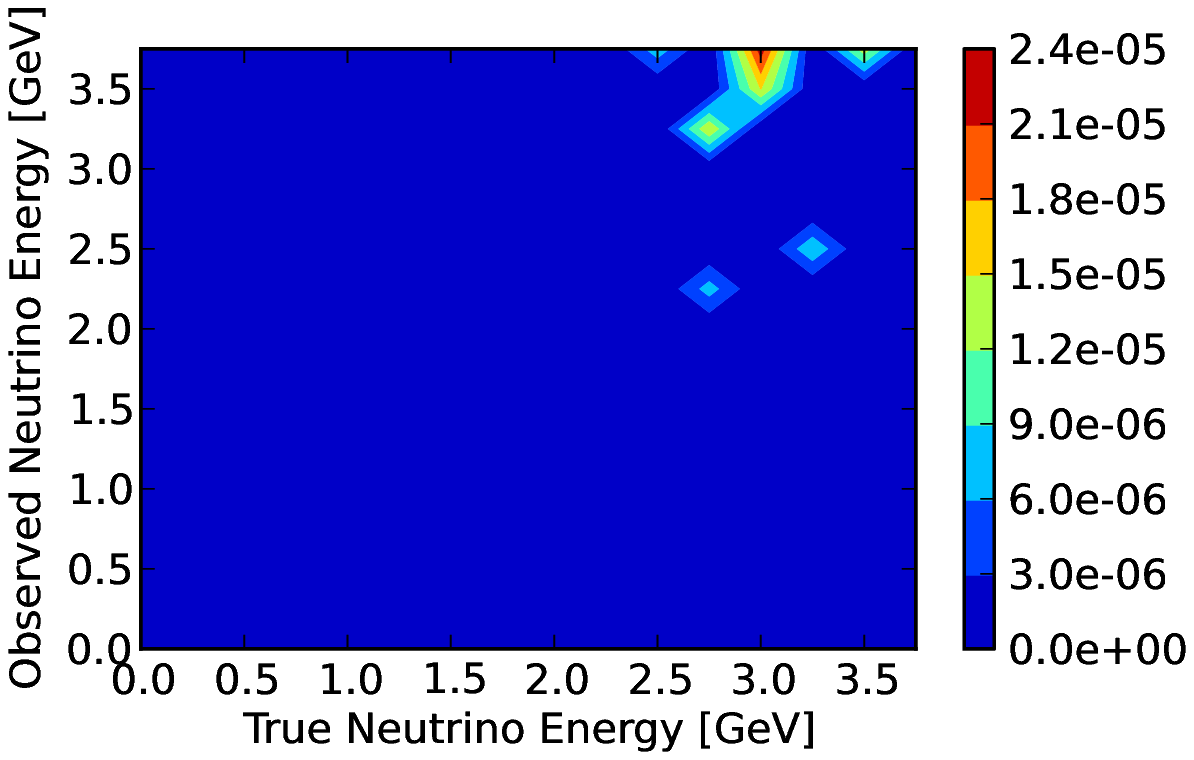}}
  \caption{Migration matrices for $\nu_\mu$ CC, $\bar{\nu}$ NC, $\bar{\nu}_\mu$ CC, and $\nu_e$ CC.}
  \label{fig:mms}
\end{figure}

Migration matrices have been computed for $\nu_\mu$ CC, $\bar{\nu}_\mu$ CC, $\bar{\nu}_\mu$ NC, and $\nu_e$ CC (See \cite{c:nustorm_loi}) and can be seen in Fig.~\ref{fig:mms}.  The background level of $\nu_e$ NC events into the signal window are negligible compared with $\bar{\nu}_\mu$ NC due to the lower energies.  These numbers are derived using a GENIE and Geant4 simulation, described in the cited text, which are MC method softwares.  Statistical fluctuations exist in the migration matrices due to computational limitations.

\begin{figure}[t]
\begin{center}
\includegraphics[width=0.7\columnwidth]{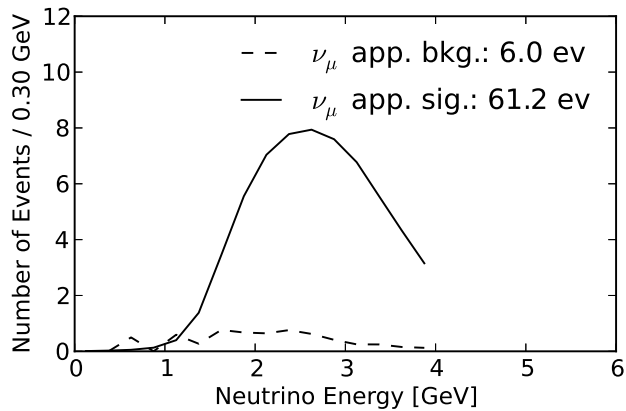}
\end{center}
\vspace*{-0.5cm}
\caption{\label{fig:rule_rate} The rule rate as a function of observed energy for the appearance channel $\nu_e \to \nu_\mu$.   Migration matrices are used for $\nu_\mu$ CC, $\bar{\nu}_\mu$ CC, $\bar{\nu}_\mu$ NC, and $\nu_e$ CC.}
\end{figure}

 \section{Statistics}
 
It is necessary to determine if the number of events observed after cuts (\emph{i.e.} rule rates) is statistically significant.  The experiment must reject the null hypothesis when accounting for statistical fluctuations.

The hypothesis $H_0$ of no oscillations is the null hypothesis and designate $H_1$ to be the alternate hypothesis.  These hypotheses have oscillation parameters associated with them: let $\theta_0 = \{ \Delta m^2_{41}, \theta_{34}, \theta_{24},\theta_{14} \}$ be the oscillation parameters associated with $H_0$, and similarly $\theta_1$ for $H_1$.   
 
The \emph{test statistic} $X$ is a function of the experimental observations and let $W$ be the space of all possible values of $X$.  One can divide $W$ into two regions: the region $w$ for those possible values of $X$ which would suggest that the null hypothesis $H_0$ is not true and the remaining region $W-w$.

It is desirable to have a small probability of $X$ -- by statistical fluctuations alone -- taking a value in $w$ when $H_0$ is true.  A level of significance $\alpha$ can be defined:

 \begin{equation}
 P(X \in w | H_0) = \alpha
 \end{equation}
 
 \noindent
 where $\alpha$ corresponds to, colloquially, ``5$\sigma$'' when $\alpha \simeq 2.8 \times 10^{-7}$ and ``10$\sigma$'' when $\alpha \simeq 7.6 \times 10^{-24}$.  The number of ``$\sigma$'' correspond to the $p$-value of having a greater than $n\sigma$ upward fluctuation of a Gaussian centered at zero.  No Gaussian assumptions are made in this analysis.
 
 The test statistic that will be used for hypothesis testing is the likelihood ratio test.    Let there be $N$ observations $\mathbf{X} = \{X_1, ..., X_N\}$ and a probability distribution function $f(X_i | \theta)$.  The likelihood function is:

\begin{eqnarray}
L(\mathbf{X} | \mathbf{\theta}) &=& \prod^N_{i=1} f(X_i | \mathbf{\theta})\\
&=&  \prod_i e^{-\lambda_i} \lambda^{X_i}_i / {X_i}!  \label{eq:like_pois}
\end{eqnarray}
 
\noindent
where $\lambda_i$ is the expected number of background in the bin with $X_i$ events and is a function of $\theta$.   The distribution is Poisson because the background levels are small.  The short baseline parameters $\theta_1$ for $H_1$ are free to take any value but the parameters $\theta_0$ are fixed to zero by the null hypothesis requiring no oscillations.  The likelihood ratio test defines a test statistic $\lambda$ such that:

 \begin{equation}
 \label{eq:lambdap}
 \lambda = \frac{L(\mathbf{X} | \mathbf{\theta_0})}{\max_{\theta_1} L(\mathbf{X} | \mathbf{\theta_1})}
 \end{equation}
 
\noindent
where the denominator is maximized with respect to $\theta_1$ while the numerator remains fixed.  Using Eq.~\ref{eq:like_pois} leads to:

 \begin{equation}
\lambda = \prod_i e^{-\lambda_i + X_i} \left( \lambda_i/X_i\right )^{X_i}.
 \end{equation}
 
The $\chi^2$ can be defined as  $\chi^2 = - 2 \ln \lambda$ (See \cite{Baker1984437}) which is preferable to using $\lambda$ because of specifics about how multiplication is performed by a computer.  Using this definition, one finds:
 
 \begin{eqnarray}
 \chi^2 = - 2 \ln \lambda = 2 \sum_i \lambda_i - X_i + X_i \ln\left(\frac{\lambda_i}{X_i}\right)
 \end{eqnarray}
 
\noindent
which has two degrees of freedom since the numerator of Eq.~\ref{eq:lambdap} has no degrees of freedom and the denominator has two degrees of freedom.
 
A test statistic has been defined that allows for determining if an experiment is sensitive to various oscillation parameters.  The $\chi^2$ can be computed in terms of energy bins, with the appropriate definition of $X_i$, allowing for spectral information to be used when computing sensitivities.

\section{The Appearance Analysis}

\begin{figure}[t]	
\begin{center}
\includegraphics[width=0.7\columnwidth]{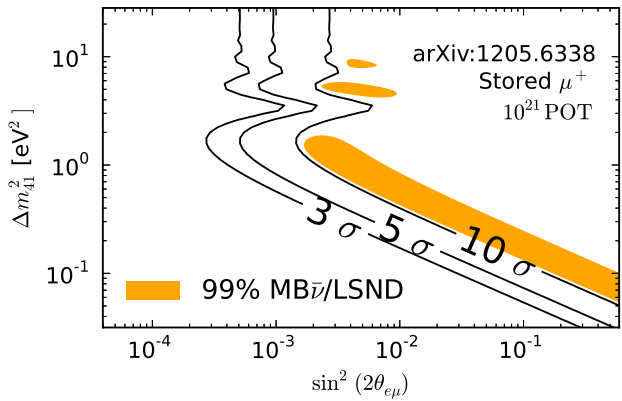}
\end{center}
\vspace*{-0.5cm}
\caption{\label{fig:contour}Sterile sensitivity under the appearance channel $\nu_e \to \nu_\mu$.  This channel is the CPT  of the LSND anomaly $\bar{\nu}_\mu \to \bar{\nu}_e$.  There is $10\sigma$ sensitivity to the LSND and MiniBooNe 99\% confidence interval \cite{Giunti:2011hn}.}
\end{figure}

The parameters to be explored in the appearance analysis are  $\Delta m^2_{41}$ and $\sin^2(\theta_{e\mu})$.  Contours in the neutrino parameter space $\Delta m^2_{41}$ versus $\sin^2(\theta_{e\mu})$ can be used to compare the sensitivities of various proposed short baseline experiments.  A statistics-only $\chi^2$ using spectral information is used (Fig.\ref{fig:contour}).

Care must be taken when defining $\chi^2(\Delta m^2_{41}, \sin^2(\theta_{e\mu}))$ to ensure that it is well-defined. In the (3+1) scenario, the signal $\nu_e \to \nu_\mu$ depends on the amplitude $\sin^2(\theta_{e\mu}) = 4 |U_{e4}|^2 |U_{\mu 4}|^2$ and frequency $\Delta m^2_{41}$  (See Eq.~\ref{eq:prob}).  If there is an appearance signal, then $|U_{e4}|^2 |U_{\mu 4}|^2 \neq 0$ which implies that both $U_{e4}$ and $U_{\mu 4}$ are nonzero.  There is disappearance of the CC and NC backgrounds (See Eq.~\ref{eq:probdisp}) which affects the background estimation in the $\chi^2$.  This issue is addressed by not oscillating the backgrounds thus overestimating the backgrounds.

\begin{figure}[t]
\begin{center}
\includegraphics[width=0.7\columnwidth]{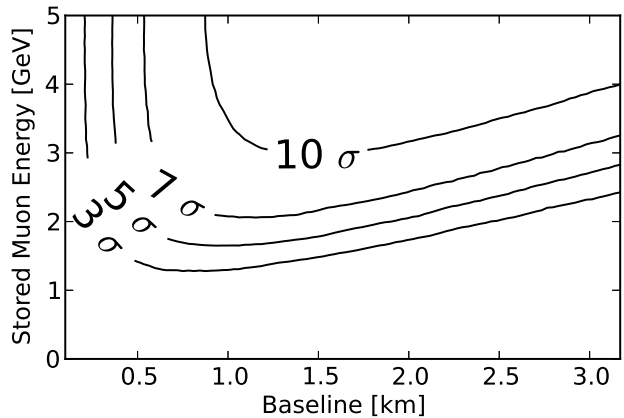}
\end{center}
\vspace*{-0.5cm}
\caption{\label{fig:sens_baseline_versus_energy} A baseline optimization using a total rates statistics-only $\chi^2$, a signal efficiency of 0.5, and background rejection of charge misidentification and NCs at $10^{-3}$ and $10^{-4}$.}
\end{figure}

\begin{figure}[t]
\begin{center}
\includegraphics[width=0.7\columnwidth]{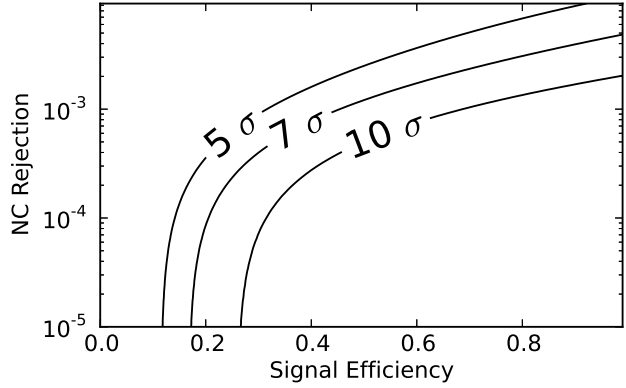}
\end{center}
\vspace*{-0.5cm}
\caption{\label{fig:sens_eff_versus_nc} Tuning the NC rejection cut.  The NC rejection level is shown versus the signal efficiency.  A charge misidentification background of $10^{-4}$ is shown to illustrate when NC backgrounds become statistically significant. A total rates statistics-only $\chi^2$ is used.}
\end{figure}

\begin{figure}[t]
\begin{center}
\includegraphics[width=0.7\columnwidth]{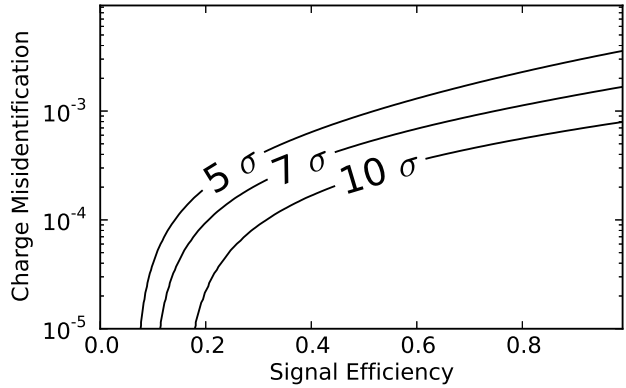}
\end{center}
\vspace*{-0.5cm}
\caption{ \label{fig:sens_eff_versus_cid} Tuning the charge misidentification cut.  The charge misidentification level is shown versus the signal efficiency.  A NC background of $10^{-4}$ is shown to illustrate when charge misidentification backgrounds become statistically significant. A total rates statistics-only $\chi^2$ is used.}
\end{figure}

\begin{figure}[t]
\begin{center}
\includegraphics[width=0.7\columnwidth]{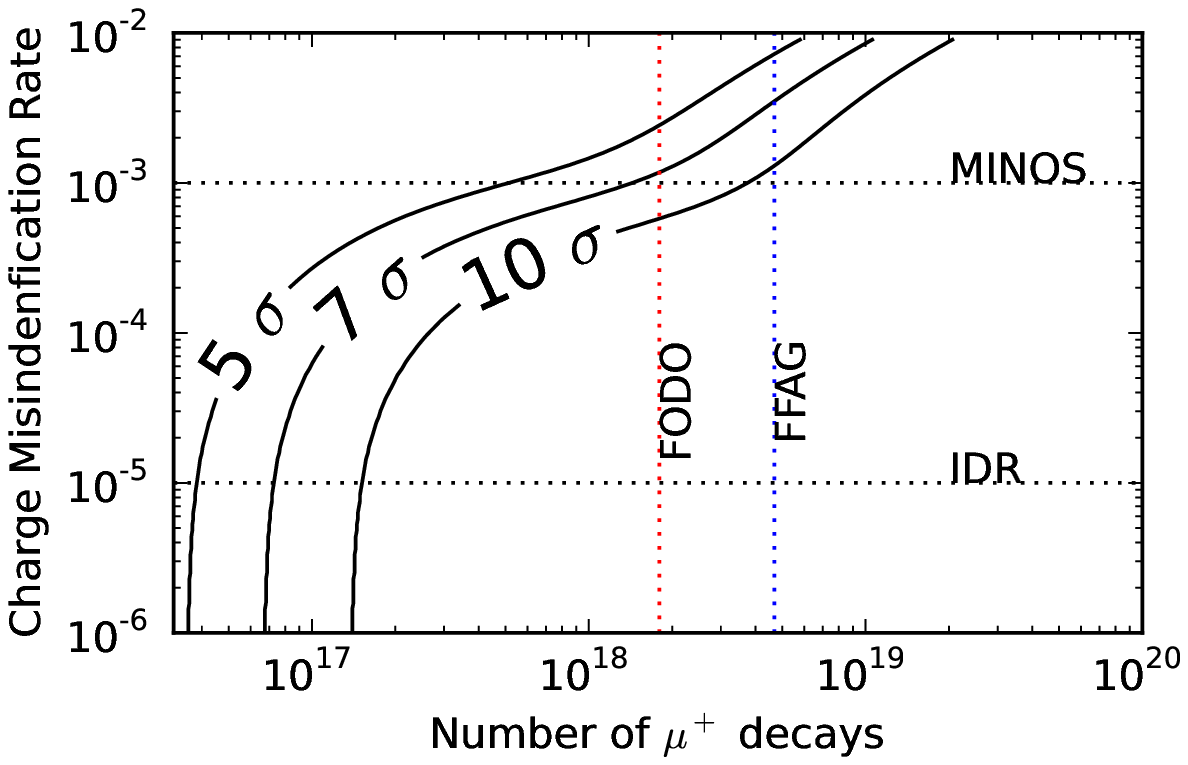}
\end{center}
\vspace*{-0.5cm}
\caption{\label{fig:sens_cid_versus_flux} An optimization between the detector performance and accelerator performance using the charge misidentification rates and number of muon decays as the performance metric.  IDR refers to the Interim Design Report \cite{NF:2011aa} detector performance.  FODO refers to the FODO lattice design that gives $1.8 \times 10^{18}$ useful muon decays whilst FFAG refers to the FFAG design that gives  $4.68 \times 10^{18}$ useful muon decays.  Both accelerators assume a front-end of the main injector at 60 GeV/c.}
\end{figure}

As the cuts-based detector performance section improves and various cost optimizations are done, there are numerous parameters that can be tuned to compensate and conserve the physics that can be done with such a facility.  For example, the optimization of baseline and energy (Fig.~\ref{fig:sens_baseline_versus_energy}) allows one to change the baseline depending on site constraints or modify the energy of the ring if the accelerator gets too expensive.  As the cuts-based detector performance improves, the various background rejections (Fig.~\ref{fig:sens_eff_versus_nc} and \ref{fig:sens_eff_versus_cid}) may allow for a smaller detector or cheaper target station.  The tools have been developed that allow the important accelerator and detector performance metrics into cost optimizations.

Figure~\ref{fig:sens_baseline_versus_energy} shows that, for a fixed baseline, increasing the muon energy is always advantageous.  This effect arises because the maximum of the $\nu_e$ flux is not at the oscillation maximum but rather at a higher energy.  At high energies the oscillation probability is:

\begin{eqnarray}
\text{Pr}[\nu_e \to \nu_\mu] &=& \sin^2 (2 \theta_{e\mu}) \sin^2 \left(\frac{\Delta m^2_{41} L}{4 E}\right)\\
&=&  \sin^2 (2 \theta_{e\mu}) \left(\frac{\Delta m^2_{41} L}{4}\right)^2 E^{-2}.
\end{eqnarray}

\noindent
The oscillation probability decreases as $E^{-2}$ for a fixed baseline.  The signal rates increase as $E^3$: there is a factor of $E^2$ from the solid angle arising from the $1/\gamma$ opening angle and another factor of $E$ from the cross section. The conclusion is that raising the stored muon energy will increase the event rates linearly with energy for a fixed baseline.  This result has been confirmed by similar analyses for other muon-decay based facilities (See sensitivity work in \cite{NF:2011aa}).

\section{Conclusion}

The sensitivity of $\nu$STORM rules out the LSND 99\% confidence interval at $10\sigma$ using only appearance information.  The appearance channel is the CPT invariant of the observed anti-neutrino LSND anomaly.  Optimizations have been shown to guide future costing and performance work.